\documentclass[twoside,leqno,twocolumn,ltexpprt]{article}  

\usepackage{amssymb}

\newtheorem{theorem}{Theorem}
\newtheorem{remark}{Remark}
\newtheorem{fact}{Fact}
\newtheorem{lemma}[theorem]{Lemma}
\newtheorem{corollary}[theorem]{Corollary}

\newenvironment{proof}{\noindent{\bf Proof}:}{\hfill $\Box$\bigskip}
\begin{document}
\def\eps{\epsilon}
\newcommand{\beq}{\begin{eqnarray*}}
\newcommand{\eeq}{\end{eqnarray*}}
\newcommand{\letab}{\le &}
\newcommand{\erf}{erf}
\newcommand{\mod}{mod}

\title{Entropy based Nearest Neighbor Search in High Dimensions }

\author{
Rina Panigrahy 
\thanks{Department of Computer Science,
Stanford University, Stanford, CA 94305.
Supported in part by Stanford Graduate Fellowship, NSF Grants EIA-0137761 and ITR-0331640, and a grant
from SNRC. {\tt rinap@cs.stanford.edu}. }
}

\maketitle \thispagestyle{empty}
\pagestyle{empty}
\begin{abstract} 
In this paper we study the problem of finding the approximate nearest neighbor of a
query point 
in the high dimensional space, focusing on the Euclidean space. 
The earlier approaches use locality-preserving hash functions (that tend to map nearby points 
to the same value) to construct several hash tables to ensure that the
query point hashes to the same bucket as its nearest neighbor in at least
one table.
Our approach is different -- we use one (or a few) hash table and hash
several
randomly chosen points in the neighborhood of the query point showing that
at least one of them will
hash to the bucket containing its nearest neighbor.
We show that the number of randomly chosen points in the neighborhood of the 
query point $q$ required depends on the entropy of the hash value $h(p)$ of a random point $p$
at the same distance  from $q$ at its nearest neighbor, given $q$ and the locality preserving hash function $h$
chosen randomly from the hash family.
Precisely, we show that if the entropy $I(h(p)|q,h) = M$ and 
$g$ is a bound on the probability that
two far-off points will hash to the same bucket,
then we can find the approximate nearest neighbor in $O(n^\rho)$ time and near
linear $\tilde O(n)$ space  where $\rho = M/\log(1/g)$. 
Alternatively we can build a data structure of size
$\tilde O(n^{1/(1-\rho)})$ to answer queries in $\tilde O(d)$ time.
By applying this analysis to the locality preserving hash functions in \cite{IM98, KOR98, DIIM04}
and adjusting the parameters we show that the $c$ nearest neighbor can be
computed in time $\tilde O(n^\rho)$ and near linear space where $\rho \approx 2.06/c$
as $c$ becomes large. 
\end{abstract}


\section{Introduction}

In this paper we study the problem of finding the nearest neighbor of a
query point
in the high dimensional Euclidean space:
given a database of $n$ points in a $d$ dimensional space, find the nearest neighbor
of a query point. This fundamental problem arises in several applications including 
data mining, information retrieval, and image search where distinctive features
of the objects are represented as points in $\mathbb R^d$ \cite{R90, S89, CH67, DDLFH90, FSN95, F98, PPS94, DW82}.
While the exact problem seems to suffer from the ``curse of dimensionality'' (that is, either
the query time or the space requried is exponential in $d$ \cite{DL76, M93}), 
many efficient techniques have been devised for finding an approximate solution whose distance
from the query point is at most $1+\eps$ times its distance from the nearest neighbor.
 \cite{AMNSW94, K97, IM98, KOR98, H01}. The best known
algorithm
for finding an $(1+\eps)$-approximate nearest neighbor of a query point
runs in time $\tilde O(d\log n)$ using a data structure of size
$(nd)^{O(1/\eps^2)}$.
Since the exponent of the space requirement grows as $1/\eps^2$, in
practice this may be prohibitively
expensive for small $\eps$.  Indeed, since even a space complexity of
$(nd)^2$ may be too large,
perhaps it makes more sense to interpret these results as efficient,
practical
algorithms for $c$-approximate nearest neighbor
where $c$ is a constant greater than one.
Also, this is meaningful in practice as typically when we are given a
query point
we are really interested in finding
a neighbor that is much closer to the query point than the other points --
the query point (say an image) really represents the `same object' as the nearest neighbor
we expect it to `match' except that they may differ a little due to noise,
or  inherent errors in how well points represents their objects,
but it is expected to be quite far
from the other points in the database which
basically represent `different objects' from the query point.

For these parameters, Indyk and Motwani \cite{IM98} provide an algorithm
for finding the $c$-approximate nearest
neighbor in time $\tilde O(d+n^{1/c})$
using an index of size $\tilde O(n^{1+1/c})$ (while their paper states 
a query time of $\tilde O(dn^{1/c})$, if $d$ is large this can easily be 
converted to $\tilde O(d+n^{1/c})$ by dimension reduction); with a data structure of near linear size,
for the hamming space, the algorithms in \cite{IM98, KOR98} require a query time of $n^{O(\log c/c)}$.
To put this in perspective,
finding a $2$-approximate
nearest neighbor requires time $O(\sqrt n)$ and an index of size $O(n \sqrt
n)$. The exponent was improved slightly in \cite{DIIM04} for $c$ in $[1,10]$ --
instead of $1/c$ it was $\beta/c$ where $\beta$ is a constant slightly
less than $1$ for $c<10$; for example when $c=2$ they can reduce the
exponent to approximately $0.42$ implying a running time of $n^{0.42}$ and
an index of size  $n^{1.42}$. Their simulation results indicate that while 
locality sensitive hashing gives faster query time over other data structures based on
$kd$-tree, it also comes at the expense of using a lot more space. 
They work with the following decision version of the  $c$-approximate
nearest neighbor  problem: given a query point,
and a parameter $r$ for the distance to its nearest neighbor,  find
any neighbor of the query point that is that distance at most $cr$.
It is well known that the reduction to the decision version adds only
a logarithmic factor in the time and space complexity \cite{IM98, H01}.

In their formulation, they use a locality sensitive hash function that maps
points in the space to a discrete space where nearby points out likely to
get hashed
 to the same value and far off points out likely to get hashed to
different values.
Precisely, given parameter $m$ that
denotes an upper bound on the probability that two points at most $r$ apart hash to the
same bucket
and $g$ a lower bound on the probability that two points more than $cr$ apart hash to the
same bucket, they
show that such a hash function can find a c-approximate nearest
neighbor in
$\tilde O(d+n^{\rho})$ time using a data structure of size $\tilde O(n^{1+\rho})$ where
$\rho =
log(1/m)/log(1/g)$.

Their approach is to construct several hash tables to ensure that the
query point hashes to the same bucket as its nearest neighbor in at least
one table.
Our approach is different -- we use one (or a few) hash table and hash
several
randomly chosen points in the neighborhood of the query point showing that
at least one of them will
hash to the bucket containing its nearest neighbor.
We show that the number of randomly chosen points in the neighborhood of the 
query point $q$ required depends on the entropy of the hash value $h(p)$ of a random point $p$
at distance $r$ from $q$, given $q$ and the locality preserving hash function $h$
chosen randomly from the hash family.
Precisely, we show that if the entropy $I(h(p)|q,h) = M$ 
then we can find the approximate nearest neighbor in 
$\tilde O(d+n^\rho)$ time and near
linear space $\tilde O(n)$ where $\rho = M/\log(1/g)$. 
Here $I(h(p)|q,h)$ denotes the
entropy of $h(p)$ for a random point $p$ at distance $r$ from $q$
given the query point $q$ and the specific hash function $h$ from the hash family
in use.  Alternatively we can build a data structure of size
$\tilde O(n^{1/(1-\rho)})$ to answer queries in $\tilde O(d)$ time.
By applying this analysis to the locality preserving hash functions in \cite{IM98, KOR98, DIIM04}
and adjusting the parameters we show that the $c$ nearest neighbor can be
computed in time $n^\rho$ and near linear space where $\rho \approx 2.06/c$
as $c$ becomes large. For $c=2$, $\rho$ turns out to be about $n^{0.69}$.
Note that $I(h(p)|q,h)$ can be much lower than $I(h(p)|h(q))$ -- the latter 
corresponds to guessing $h(p)$ from $h(q)$ and can lead to much slower algorithms.
For example in the Euclidean case an algorithm based on the latter entropy would
give a much higher value of $\rho$ of about $\Theta(\log c/c)$, but using both
$h$ and $q$ in conjunction instead of just $h(q)$ gives us the improved results.
We also show that if the points are chosen randomly from a spherical gaussian distribution
(section \ref{secrandom}) the value of $\rho$ can be improved to about $1.47/c$

A major advantage of such a small index of size $\tilde O(n)$ is that the
entire
index
could possibly fit in main memory making all  memory accesses  RAM
accesses
instead of the much slower disk accesses. This suddenly increases the number of
possible accesses in the same query time by a factor of 1000's ! If there is a unique 
$c$-approximate nearest neighbor -- which
may be typical in practice -- we argue that  only 
$2\log n$ bits of storage are required in the index for each point
for large enough values of $c$. So even with a million entries, we 
need only an index of size 5MB which is a trivial amount of RAM space in today's PCs.

Application of our techniques to the L1 norm does not result in any improvement over
the previous results -- with linear space we get a value of 
the value of $\rho$ about $\log (c)/c$ matching the bounds in \cite{KOR98, IM98}.

\section{Results}

\begin {itemize}

\item $B(p,r)$:
Let $B(p,r)$ denote the sphere of radius $r$ centered at $p$ a point in
$\mathbb R^d$; that is the set of points at distance $r$ from $p$.

\item $I(X)$:
For a discrete random variable $X$, let $I(X)$ denote its information-entropy.  For
example if
$X$ takes $N$ possible values with probabilities $w_1, w_2, ..., w_N$ then
$I(X) = I(w_1, w_2, .., w_N) = \sum I(w_i) = \sum -w_i \log w_i$

\end {itemize}

We will work with the following decision version of the  $c$-approximate
nearest neighbor  problem: given a query point
and a parameter $r$ indicating the distance to its nearest neighbor,  find
any neighbor of the query point that is that distance at most $cr$.
We will refer to this decision version as the $(r, cr)$-nearest neighbor
problem and a solution to this as a $(r, cr)$-nearest neighbor.
It is well known that the reduction to the decision version adds only
a logarithmic factor in the time and space complexity \cite{IM98, H01}.

We use locality preserving hash functions to map database points into
a hash table; a locality preserving hash function is a random function from a 
hash family that is {\em likely} to hash nearby points to the same value and
far off points to different values in a discrete space.
To find the approximate nearest neighbor of a query point, we hash several randomly chosen points in the vicinity
of the query point and show that the approximate nearest neighbor is
likely to be present in one of these buckets.

We assume that the locality preserving hash function has the following
properties. Let $M$ denote the entropy $I(h(p)|q,h)$ where $p$
is a random point in $B(q,r)$. Here $I(h(p)|q,h)$ denotes the
entropy of $h(p)$ given the query point and the specific hash function from the
hash family in use. Let $g$ denote an upper bound on the probability that
two points that are at least distance $cr$ apart will hash to the same
bucket. Note that after a random rotation and a random shift of the origin
 the nearest neighbor of $q$ appears like a random point on $B(q,r)$. 
Our algorithm is simple:

{\bf Construction of hash table:}
Pick $k = \log n/\log(1/g)$ random hash functions $h_1, h_2, .., h_k$.
For each point $p$ in the database compute (after random rotations and shifts
for each hash function) $H(p) = (h_1(p), h_2(p), ..,
h_k(p))$
For each point $p$, store it in a table at location $H(p)$; use hashing to store only the
the nonempty locations. Use $polylog n$ such randomly constructed hash tables.

{ \bf Search Algorithm:} To find a point at distance at most $cr$ from a query
point $q$ given
that there is a neighbor at distance at  most $r$ from $q$, 
pick $\tilde O(n^\rho)$ random points $v$ from $B(q,r)$ and search in the
buckets $H(v)$. Here $\rho = M/\log(1/g)$.

\begin{theorem}\label{mainthm}
With probability at least $\tilde O(1)$, if the nearest neighbor of the query
point is at distance $r$, the search algorithm finds a neighbor at
distance at most $cr$.
With constant probability, no more than $\tilde O(n^\rho)$ time is spent
searching points
that are at a distance more than $cr$ from $q$.
\end{theorem}

By using $polylog n$ hash tables our algorithms can be made to succeed with
high probability.

Alternatively, we show that our methods can be used to construct a data structure of size 
$\tilde O(n^{1/(1-\rho)})$ to answer queries in $\tilde O(d)$ time.

By applying this analysis to the locality preserving hash functions from \cite{KOR98, IM98, DIIM04}
and adjusting the parameters we show that the $c$ nearest neighbor can be
computed in time $\tilde O(n^\rho)$ and near linear space where $\rho \approx 2.09/c$
as $c$ becomes large. For $c=2$, $\rho$ turns out to be about $0.69$.

We start in section \ref{secprelim} with preliminaries including a crucial lemma
that states the number of random samples required for an arbitrary random variable to guess
its specific value. To simplify the exposition of the basic principles, 
in section \ref{secrandom} we study a random instance of the
nearest neighbor problem in Euclidean space where the points in the database are 
chosen randomly from a spherical gaussian distribution. In section \ref{secgeneral} we prove the main theorems
applicable to nearest neighbor search for arbitrary point sets and derive algorithms for
nearest neighbor search in Euclidean space. Finally, in section \ref{secimplementation}
we discuss some computational issues relevant for practical implementation.


\section{Preliminaries}\label{secprelim}

First let us go through some notations.

\begin {itemize}

\item $N(\mu,r), \eta(x)$:
Let $N(\mu,r)$ denote the normal distribution with mean $\mu$ and variance $r^2$
with probability density function given by $\frac{1}{r\sqrt{2\pi}}
e^{-(x-\mu)^2/(2r^2)}$. Let $\eta(x)$ denote the function $\frac{1}{\sqrt{2\pi}}e^{-x^2/2}$.

\item $N^d(p,r)$:
For the $d$-dimensional Euclidean space, for a point $p = (p_1, p_2, ..., p_d) \in
\mathbb R^d$
let $N^d(p,r)$ denote the normal distribution in
$\mathbb R^d$ around the point $p$ where the $i$th coordinate of a random point
has the normal distribution
$N(p_i, r/\sqrt d)$ with mean $p_i$ and variance $r^2/d$. 
It is well known that this distribution is spherically symmetric around $p$. A point from
this distribution is expected to be at
root-mean squared distance $r$ from $p$; in fact, for large $d$ its distance
 from $p$ is close to $r$ with high probability (see for example
lemma 6 in \cite{IM98})

\item $\erf(x), \Phi(x)$:
The well-known error function $\erf(x) = \frac{2}{\sqrt \pi}\int_0^x e^{-x^2} \,dx$, is equal
to the probability that
a random variable from $N(0,1/\sqrt 2)$ lies between $-x$ and $x$.
 Let $\Phi(x) = \frac{1 - erf(x/\sqrt 2)}{2}$.
For $x \ge 0$, $\Phi(x)$ is the probability that a random variable from 
the distribution $N(0,1)$ is greater than $x$.

\item
Use $\alpha \approx 1.303$ to denote the constant: $\int_0^\infty
I(\Phi(x),1-\Phi(x)) \,dx$. The approximate value of this integral 
has been computed using Matlab.

\item Projection:
We will use the following commonly used projections that map points in
Euclidean space to real numbers. Let $v$ denote a random vector from the
distribution
$N^d(0, \sqrt d)$.
Then for any point $p \in \mathbb R^d$, the projection $f(p) = v.p$ is distributed
according to
the normal distribution $N(0, ||p||)$ where $||p||$ is the Euclidean distance of $p$ from the origin.
Several such projections can be used to project a point $p$ into a low
(say $k$) dimensional space -- for example, we can have the function
$F(p) = (f_1(p), f_2(p), ..., f_k(p))$ for random choices of projection
functions $f_1, ..., f_k$.

\end{itemize}

The following are well known facts about such random  projections (they are direct consequences of
the $2$-stability of the normal distribution \cite{Z86}):

\begin{fact}
Under a random projection described above, for any points $p$ and $q$, $F(p)-F(q)$ has
the distribution $N^k(0, d(p,q))$ where $d(p,q)$ denotes the distance between $p$ and $q$. 
So the distribution of $F(p)-F(q)$
depends only on the distance $d(p,q)$ and not on the positions of $p$ and $q$. 
\end{fact}

\begin{fact}
If $r$ is random point on $B(p,r)$, then $F(r)-F(q)$ has the distribution $N(0, \sqrt (d(p,q)^2 + x^2))$.
\end{fact}

{\it Guessing the value of a random variable:}
If a random variable takes one of $N$ discrete values with equal probability
then a simple coupon collection based argument shows that if we guess
$N$ random values at least one of them should hit the correct value with constant 
probability. The following lemma states the required number of samples
for arbitrary random variables so as to `hit' a given random value of the
variable. It essentially states how many guesses are required to guess
the value of a random variable.

\begin{lemma}\label{couponcoll}
Given an random instance $x$ of a discrete random variable with a certain
distribution $\Omega$ with entropy $I$, if $O(2^I)$ random samples are chosen
from this distribution at least one of them is equal to
$x$ with probability at least $\Omega(1/I)$.
\end{lemma}
\begin{proof}
Let $w_1, w_2, ..., w_N$ denote the probability distribution $\Omega$ of the
discrete space.

After $s = 4.(2^I+1)$ samples the probability that $x$ is chosen is $\sum_i
w_i [1-(1-w_i)^s]$.

If $w_i \ge 1/s$ then the term in the sum is at least $w_i(1-1/e)$.
So if all the $w_i's$ that are at least $1/s$ add up to at least $1/I$ then
the above sum is at least $\Omega(1/I)$. Otherwise we have a collection of
$w_i's$ each of which is at most $1/s$ and they together add up to more
than $1-1/I$.

But then by paying attention to these probabilities we see that
the entropy $I = \sum_i w_i \log (1/w_i) \ge \sum_i w_i \log s \ge (1-1/I)\log s
                \ge (1-1/I)(I+2) = I + 1 - 2/I$. For $I \ge 4$, this is strictly greater than $I$,
 which is a contradiction. 
If $I < 4$ then the largest $w_i$ must be at least $1/16$ as otherwise a similar argument shows that 
$I = \sum_i w_i \log (1/w_i) > w_i \log 16 = 4$, a contradiction; so in this case even one sample 
guesses $x$ with constant probability.
\end{proof}

\begin{remark}\label{diffdistribution}
While the above lemma assumes that the random samples are chosen from the same 
distribution from which $x$ was derived, it is easy to extend it to the case 
where random samples are chosen from a distribution slightly different from 
$\Omega$, where say the probabilities of corresponding events differ at most by 
a constant factor. For example the random samples could be chosen from a 
distribution $\Omega' = (w_1', w_2', ..., w_N')$ where the individual probabilities 
differ from the ones in the distribution $\Omega = (w_1, w_2, ..., w_N)$ by at most 
a constant multiplicative factor. 
\end{remark}

\begin{remark}
The above result is tight to the extent that you cannot get a probability
much better than
$\Omega(1/I)$ with  $O(2^I)$ samples. There is a distribution with entropy
$I$ so that even
picking $O(I2^I)$ samples will hit $x$ only with probability $O(\log
I/I)$. The distribution
has one element with probability $4\log I / I$ and all others with equal
probability of $O(1/(I^2 2^I))$.
The converse of the lemma is not necessarily true. That is, there may be a
distribution with entropy $I$, 
and it may be sufficient to pick much fewer than $2^I$ samples –- in fact
just one sample -- to hit $x$
is significant probability. Think of a distribution where one element has
probability $1/2$ there are
an exponentially large number of remaining elements with tiny uniform
probability.
\end{remark}

\section{Random Instance in Euclidean Space}\label{secrandom}

We study a random instance of the problem where each point is distributed
according to $N^d(0,1/\sqrt 2)$. The reason we choose this distribution with
a deviation of $1/\sqrt 2$ is because the expected distance between any two 
points is $1$; in fact, the distance is very close to $1$ with high probability
for large $d$. 
The query point is randomly chosen around a certain point p with
distribution $N^d(0,1/c)$; the query point is at distance close to
$1/c$ from its nearest neighbor.
The idea is to use the random projections to a real line introduced earlier.

For two points separated by distance $x$, the distance in the projection
is distributed as $N(0,x)$.
We use $k=\log n$ such projections. For each point $p$ this gives a vector of
real numbers
$F(p) = (f_1(p), f_2(p), ..., f_k(p))$.
For each projection we produce a bit $h_i(p)=0$ if $f_i(p)<0$ and $1$
otherwise, 
giving $H(p) = (h_1(p), h_2(p), ..., h_k(p))$
This hashes each point to an element of $\{0,1\}^k$.
If $k=\log n$, the number of points in any one hash bucket (bin) is at most
$\log n$ with high probability.

Unfortunately, the query point $q$ may not hash to the same bucket as it's
nearest neighbor $p$.
We will try to guess $H(p)$. It can be shown that the hash values $H(p)$ and $H(q)$ are expected to differ
in about $O(1/c)$ fraction of the bits. Based on this fact
we may need to search a large number of hash buckets, up to ${k \choose
k/c} \approx n^{I(1/c, 1-1/c)}
\approx n^{O(\log c/c)}$ for large $c$.

Our essential observation is that this search space can be pruned
significantly by paying attention to the
vector $F(q)$ from which $H(q)$ is derived.
If a coordinate $f_i(q)$ is far from $0$, it is less likely that $h_i(p)$ and
$h_i(q)$ will differ. In fact,
if the absolute value,  $|f_i(q)| = x$ then for $h_i(p)$ and $h_i(q)$ to differ, the projection
$f_i$
must map $p$ at least $x$
away from $q.$ This happens with probability at most $e^{-O(x^2c^2)}$. This is
exponentially small in $c$ except
when $x$ is comparable to $1/c$ which happens only with probability about
$1/c$.
So the search space of $H(p)$ given $F(q)$ is much smaller than ${k
\choose k/c}$.
To estimate the size of this search space precisely we compute the entropy of
$H(p)$ given $F(q)$.
If this is $M$, then by lemma \ref{couponcoll} the search space is
about $O(2^M)$.

Now, $I(H(p)|F(q) \le \sum I(h_i(p)|f_i(q))$.
Let us first compute $I(h(q)|f(p))$ for one random projection.

\begin{lemma}\label{entropyrandom}
If $p$ is a random point on $B(0,1/\sqrt 2)$ and $q$ is a random point on
$B(p,1/c)$,
then for a random projection
$I(h(q)|f(p)) = \frac{1}{c} (1-o(1)) 2\alpha/\sqrt \pi  \approx 1.47/c$
\end{lemma}
\begin{proof}
f(p) has the distribution $N(0,1)$, and $f(q)-f(p)$ has the distribution
$N(0,1/c)$.
So $f(p)$ is at distance $x$ from $0$ with probability density $2\eta(\sqrt 2 x)$ and
in that
case the probability that $f(q)$ is not on the same side as $f(q)$ is
$\Phi(xc)$, so
the entropy of $h(q)$ is $I(\Phi(cx), 1-\Phi(cx))$.
So

\beq
I(h(q)|f(p)) &=& \int_0^\infty 2 \eta(\sqrt 2 x) I(\Phi(cx), 1-\Phi(cx)) \,dx \\
        &=&   \frac{2}{\pi} \int_0^\infty e^{-x^2} I(\Phi(cx), 1-\Phi(cx)) \,dx \\
        &=&  \frac{2}{c\pi} \int_0^\infty e^{-(x/c)^2)} I(\Phi(x), 1-\Phi(x))
\,dx
\eeq

Now $\Phi(x) \le e^{-x^2/2}/x$ drops exponentially and for large $c$,
$e^{-(x/c)^2}$ drops slowly and is close to $1$ until $x$ becomes
comparable to $c$.
So
$ \int_0^\infty e^{-(x/c)^2} I(\Phi(x), 1-\Phi(x))\,dx =
(1-o(1))\int_0^\infty I(\Phi(x), 1-\Phi(x)) \,dx$

\end{proof}

Similarly it can be shown that $I(h(p)|f(q)) = I(h(q)|f(p)) \approx 1.47/c$ (see appendix \ref{entropyqp}).
So $I(H(p)|F(q)) \le 1.47 k/c$.  But $I(H(p)|F(q))$ is the expected entropy of $H(p)$ given 
$F(q)$ for random choices of $q$ from $B(p,1/c)$. We will argue that for large $d$, even for a
fixed random choices of $q$ and $f$, $I(H(p)|F(q)) \le (1+o(1))1.47 k/c$ with high probability
of $1-o(1)$: 
Observe that if $d \ge k$ the tuples $(f_i(q), f_i(p))$ are independent for the $k$
different values of $i$; so the sum $\sum I(h_i(p)|f_i(q))$ is a sum of independent random
variables in the range $[0,1]$ each with expectation $1.47/c$. By chernoff bounds, with high probability the sum will be close
to the mean. Even if $d < k$ the terms are $d$-wise independent and chernoff bounds may be
applied to $d$ terms at a time; the high probability bound follows if we assume $d$ is large.
This means by lemma \ref{couponcoll}, with high probability of $1-o(1)$, 
the search time is about $2^{(1+o(1))1.47k/c}$ which is $n^{(1+o(1))1.47/c}$.

The algorithm is as follows:
For $n^{(1+o(1))1.47/c}$ iterations:
Search a random bucket from the distribution of $H(p)$ given $F(q)$.
Report the nearest neighbor among all points searched.

Note that $F(p)$ given $F(q)$ has a normal distribution (appendix \ref{entropyqp}) 
and so sampling
with the same distribution as $H(p)$ given $F(q)$ is easy. This gives an algorithm that takes near linear space and  
$n^{(1+o(1))1.47/c}$ time.

\begin{remark}
In the decision version of the nearest neighbor problem we assumed 
that we know the exact distance $1/c$ to the nearest neighbor whereas 
in earlier works, $1/c$ is only an upper bound on the
distance to the nearest neighbor. This can easily be fixed by guessing 
the exact distance within a factor of $1+\eps$ where $\eps=O(1/\log n)$. 
So $H(p)$ has almost the same probability distribution as the nearest 
neighbor of $q$.
Then it follows from remark \ref{diffdistribution} that we can still apply
lemma \ref{couponcoll} to achieve the same result. The search time only 
increases by a factor of $O(\log n)$.
\end{remark}

\begin{remark}
In our search data structure, we have used a set of $\log n$ random hyperplanes 
to separate the $n$ points of the database. It can be shown that if the points 
can be separated by `thick' hyperplanes -- say $\log n$ almost orthogonal hyperplanes
of thickness at least $t$, then $I(h(p)|q,h) = e^{-O(c^2/t^2)}$ implying a much
faster search time of $n^{e^{-O(c^2/t^2)}}$ if $t$ is not too large.
While such thick hyperplanes exist for large dimensions when $d \ge n$ (see appendix \ref{thickhyper}),
for $d << n$ a simple probabilistic calculation shows that such a set of thick
separating hyperplanes does not exist.  
\end{remark}

Note that for large $d$, we need not store the entire description of each point
in the hash table but only its $O(\log n)$ bit hash value. With high probability, 
this should be sufficient to distinguish between points that are $1/c$ close to the 
query point from points that are at least $1$ away. 

Alternatively, we will show later in section \ref{secgeneral} how this
technique can also be used to search in $\tilde  O(d)$
time and $n^{(1+o(1))/(1-1.47/c)}$ space.

Although we have assumed that the points are chosen randomly from a normal distribution, our
results in this section can be applied to any set of points whose pairwise
distances are about the same. This is true when points are chosen randomly 
from other distributions such as from a cube.
In that case we can set the origin to be the centroid of the point set.
It can easily be shown
 that the distance of any point from the centroid is about $1/\sqrt 2$ of
the interpoint distance.

\section{Generalizing to arbitrary set of points}\label{secgeneral}
\subsection{Proof of Main theorems}

We now generalize our techniques to arbitrary set of points. Assume without 
loss of generality that the nearest neighbor of the query point is at distance
$1/c$ from the query point, and we are interested in finding any point at distance 
at most $1$ from the query point. 
We use locality preserving hash functions to map database points into
a hash table.
To find the approximate nearest neighbor of a query point, we hash several randomly chosen points in the
$1/c$-neighborhood of the query point and show that a $(1/c,1)$-nearest neighbor is
likely to be present in one of these buckets.
Let $M$ denote the entropy $I(h(p)|q,h) = M$ where $p$
is a random point in $B(q,r)$. Here $I(h(p)|q,h)$ denotes the
entropy of $h(p)$ given the query point and the specific hash function from the
hash family in use. Let $g$ denote an upper bound on the probability that
two points that are at least distance $1$ apart will hash to the same
bucket.
Pick $k = \log n/\log(1/g)$ random hash functions $h_1, h_2, .., h_k$
(after random rotations and shifts) and store each point $p$ in the database
in the bucket $H(p) = (h_1(p), h_2(p), ..,
h_k(p))$. Since many buckets may be empty we use hashing to only store the non-empty buckets.

First observe that after a random rotation and a random shift of the origin,
 the nearest neighbor of $q$
appears like a random point  $p$ on $B(q,1/c)$ (this rotation and shift may not be required as
the hash functions may already perform them implicitly, see section \ref{subseclsh}). 
We will show how to guess $H(p)$ in
time $\tilde O(n^{M(1+1/\log n)/\log g})$. Since we are only interested in
running times where the exponent of $n$ is at most $1$, this is $\tilde
O(n^{M/\log g})$.

Now $I(H(p)|q, H) \le \sum_1^k I(h_i(p)/q, h_i) = kM$. This means on an average at most 
$kM$ bits are required to guess $H(p)$ for a
given set $H$ of $k$ random hash functions.
$I(H(p)|q, H)$ also denotes the expected value of
$I(H(p)|q)$ under random choices for fixing the set $H$ of hash functions. So for a fixed $H$,
by Markov inequality with at least probability $1/\log n$, this entropy is at most
$kM(1+1/\log n)$.
Let us
assume this is the case.
We are now ready to prove theorem
\ref{mainthm}.\\

\begin{proof} {\bf [of theorem \ref{mainthm}]}
For a given set $H$ of $k$ hash functions lemma \ref{couponcoll} implies that by picking $2^{kM(1+1/\log n)}$ random
values with the same distribution as $H(p)$, at least one of them is equal
to
$H(p)$ with at least $O(1/(kM))$ probability. So with one hash table with probability
at least $O(1/(kM))$, we can find $H(p)$ in time $2^{kM(1+1/\log n)}$.
Also picking random variables with the distribution as $H(p)$ is easy:
just compute $H(r)$ where $r$ is a random point from $B(q,1/c)$.
So by lemma \ref{couponcoll}, by searching $O(2^{kM(1+1/\log n)})$
buckets obtained by applying
$H$ on randomly chosen points $v$ in $B(q,1/c)$,
with probability at least $O(1/(kM))$ we find the nearest neighbor $p$.
Setting $k = \log n/\log (1/g)$ gives us the desired result.

We also need to bound the number of far off points visited over the
$O(2^{kM(1+1/\log n)})$ buckets searched. For any point $t$ that is at least
distance $1$ from $q$,
the probability that it is visited in one bucket
is at most $g^k$.  So out of $n$ such possible points the expected
number of such points visited over all buckets is at most $ng^k O(2^{kM(1+1/\log n)}) =
O(2^{kM(1+1/\log n)})$. So with probability $1/2$ at most twice
 as many far off points are visited.

\end{proof}

So in the end the algorithm is simple:
Pick $O(2^{kM(1+1/\log n)})$ random points from $B(q,1/c)$. 
Search the buckets these points hash to, limiting the total number of points visited at
distance more than $1$ from $q$ to at most $O(2^{kM(1+1/\log n)})$.
Repeat this for $polylog n$ hash tables and pick the nearest found neighbor.

Alternatively by storing p in buckets obtained by applying $H$ on $2^{kM(1+\eps)}$
randomly chosen points from $B(p,1/c)$,
we can have a small search time with slightly more space. For a fixed random choice of $H$, 
by Markov's inequality
the probability that $I(H(q)|p)$ exceeds $kM(1+\eps)$ is at most $\eps$.
By lemma \ref{couponcoll} with
probability at least $O(\frac{1}{kM(1+\eps)})$ the query point will hash to one of
these buckets.
Again how many far off points can be present in this bucket? A given point
$t$
in the database that is at distance at least $1$ away from $q$  will be stored in
$2^{kM(1+\eps)}$ buckets. These buckets are $H(v)$ for
$O(2^{kM(1+\eps)})$
randomly chosen values $v$ picked from $B(t,1/c)$. Again if $g$ denotes an
upper bound on the  probability that one such random point $v$ hashes to the
same bucket as $q$,
then the probability that for one
such $v$, $H(v)=H(q)$ is at most $g^k$.


 So over $O(2^{kM(1+\eps)})$ choices of
$v$ from $B(t,1/c)$,
the probability that any of these hash to the same bucket as $q$ is at most
$2^{kM(1+\eps)} g^k$. Out of the $n$ points in the database the
expected number
of points that hash to the same bucket as $q$ is at most $n 2^{kM(1+\eps)}
g^k$. We
choose $k$ so that this is at most $1$, giving $k = \log n /(\log(1/g)-M(1+\eps))$.
 So by Markov's inequality the probability
that more
than $2$ points distance at least $1$ from $q$ hash to the same bucket as
$q$ is
at most $1/2$. Again by using $O(\log n)$ hash tables with high
probability at least
for one of them not more than $2$ far off points will be searched. We
limit the search
in each bucket to at most $3$ points. Here the size of the hash table is $O(n 2^{kM(1+\eps)})
= O(n^{\log(1/g)/(\log(1/g)-M(1+\eps))}) = O(n^{1/(1-\rho(1+\eps))})$. This is $O(n^{1/(1-\rho)})$ if 
$\eps = (1-\rho)^2/\log n$. The total success probability is $O(\frac{\eps}{kM(1+\eps)}) = \tilde O((1-\rho)^3)$

So we have proved the following theorem.

\begin{theorem}\label{mainthm2}
With probability at least $\tilde O((1-\rho)^3)$ if we use $k = \log n /(\log
(1/g)-M(1+\eps))$
projections, using a hash table of size $O(n^{1/(1-\rho)})$ the
search algorithm succeeds for
one hash table.
With constant probability, no more than $\tilde O(1)$ points
that are at a distance more than $1$ from $q$ are searched.
\end{theorem}

Again, by using $polylog n$ hash tables the algorithm can be made to succeed with
high probability.

\subsection{Choice of Hash functions for Euclidean Space}\label{subseclsh}

We now apply our techniques on the locality preserving hash functions
for Euclidean space \cite{KOR98, IM98, DIIM04}.

Instead of mapping $f(p)$ to a bit we map it to an integer. As in \cite{
IM98, DIIM04}, divide the real line into
equal sized intervals of size $D$ and add a random shift. 
Precisely, the point $p$ is hashed to an integer 
 $h(p) = \lfloor (f(p) + \beta)/D \rfloor = \lfloor (p.v + \beta)/D
\rfloor $
where $v$ is a random vector from the distribution $N^d(0,\sqrt d)$ and
$\beta$ is a
random number in $[0,D]$.  $H(p) = (h_1(p), h_2(p), ..., h_k(p))$.
Essentially $H$ divides maps the space $R^k$ into a grid of cubes of side
length $D$.

Let $r_i(p) = (f_i(p)+ \beta_i) \mod D$. So $R(p) = (r_1(p), r_2(p), ..., r_k(p))$
denotes
the relative position of F(p) within its cube. $R(p)$ is uniformly
distributed
in $[0,D]^k$. We will later set $D$ to be about $3$.

Now consider two points $p$ and $q$ that are distance $1/c$ apart. We will
try to guess the relative position of $p$'s subcube $H(p)$ from $q$'s
subcube $H(q)$, given the position $R(q)$ of $q$ in its subcube;
that is we will try to guess $H(p)-H(q)$.
Under the $k$ random projections, $F(p) - F(q)$ is randomly distributed
according
to $N(0,1/c)^k$ and is independent of the relative position $R(q)$ in its
cube
as the alignments of the intervals are independent of the projections $f_i$.
Time required to guess $H(p)-H(q)$ depends on the entropy of $H(p)-H(q)$
given $R(q)$.

The following lemma computes $I(h_i(p)-h_i(q)|r_i(q))$

\begin{lemma}\label{entropy}
If $p$ and $q$ are distance $1/c$ apart then under a random projection,
$I(h(p)-h(q)|r(q)) =  \frac{1}{c}(1+e^{-O(c^2D^2)})   2\alpha/D$
where $\alpha = \int_0^\infty I(\Phi(x),1-\Phi(x)) \,dx$
\end{lemma}
\begin{proof}
r(p) is a random value in $[0,D]$; so the probability density that it takes
value $x$ in $[0,D]$ is $1/D$. $h(p)-h(q)$ takes integral values, however,
as $c$ becomes large, in
terms of its entropy most of it is concentrated at $1$ and $-1$.
Let $M_i$ denote $I(h(p)-h(q) = i|r(p))$.  We are interested in the sum
$\sum_i M_i$
over all integers $i$. By symmetry $M_i=M_{-i}$.
If $r(p)=D-x$,
$Pr[h(p)-h(q) = 1] = \Phi(cx)-\Phi(cx+cD)$

So 
\beq
M_1 &=& I(h(p)-h(q) = 1|r(p))\\ 
&=& \frac{1}{D} \int_0^D I( \Phi(cx)- \Phi(cx+cD))\,dx \\
&=& \frac{1}{cD} \int_0^{Dc} I( \Phi(x)- \Phi(x+cD)) \,dx
\eeq

Again as $\Phi(x)$ drops exponentially, $\Phi(x+cD)$ is
negligible as compared to $\Phi(x)$,
and further the integral to $\infty$ is not much more as than the integral
to $Dc$.
So $\int_0^{Dc} I( \Phi(x)- \Phi(x+cD))  \,dx
= (1 - e^{-O(c^2D^2)})\int_0^{\infty} I(\Phi(x)) \,dx$

We have shown that
$M_1$ (and $M_{-1}) = (1 - e^{-O(c^2D^2)})\frac{1}{cD} \int_0^{\infty} I(\Phi(x)) \,dx$.
Also $M_i$ drops exponentially with $i$ since for a given value of $r(p)$,
$Pr[h(p)-h(q) = i]$ drops exponentially with a factor of $e^{-(Dc)^2/2}$.

\beq
M_0 &=&  \frac{2}{D} \int_{0}^{D/2} I( 1-\Phi(cx)-\Phi(Dc-xc) ) \,dx \\
&=&  \frac{2}{cD} \int_{0}^{Dc/2} I( 1-\Phi(x)-\Phi(Dc-x) ) \,dx
\eeq

Again as before we argue that in the range $[0,Dc/4]$,
$\Phi(Dc-x)$
is negligible as compared to $\Phi(x)$, and beyond that
they are both negligible $(e^{-O(c^2D^2)})$.
This gives us,
\beq
M_0 = (1+e^{-O(c^2D^2)}). \frac{2}{cD} \int_{0}^{\infty} I(1-\Phi(x)) \,dx
\eeq

So, $\sum_i M_i = (1+e^{-O(c^2D^2)})(M_{-1}+M_0+M_1) = 
		(1+e^{-O(c^2D^2)}) \frac{2}{cD} \int_0^\infty I(\Phi(x),1-\Phi(x)) \,dx$

\end{proof}

The following lemma computes the probability $g$ that a point $t$ at distance at
least $1$ from $q$ hashes to the same
value $h(t)$ as $h(q)$ under one projection.

\begin{lemma}\label{gvalue}
$g = 1 - \frac{1}{D} \sqrt{\frac{2}{\pi}} (1-e^{-D^2/2})$
\end{lemma}
\begin{proof}
If $f(t)$ and $f(q)$ are $x$ apart then the probability that they are
separated by the interval boundaries is $x/D$.
A simple computation shows that the probability $1-g$ that $t$ and $q$ hash to
different values in one projection is
$2\int_0^D (x/D)\eta(x) \,dx = \frac{1}{D} \sqrt{\frac{2}{\pi}}(1-e^{-D^2/2})$
\end{proof}

Now since the function $r$ is implicit in the description of the function $h$,
$I(h(p)|q,h) \le I(h(p)-h(q)/r(q))$. So by theorem \ref{mainthm} we have:

\begin{corollary}\label{rhovalue}
A $c$-approximate nearest neighbor in the Euclidean space
can be found in time $\tilde O(n^{\rho})$ using a data structure of size 
$\tilde O(n)$  where
$\rho = 2\alpha/[D log(1 - \frac{1}{D} \sqrt{\frac{2}{\pi}}
(1-e^{-D^2/2}))]$
\end{corollary}

Setting $D=3$ gives the value of $\rho = 2 \alpha /1.26 \approx 2.06$

Alternatively, using a data structure of size $\tilde O(n^{1/(1-\rho)})$, we  can
perform the search operation in $\tilde O(d)$ time: again, if $d(t,q)>1$, the
upper bound of $g$ still holds on the probability that a random point in
$B(t,1/c)$ hashes to the same value as $q$.
This is because under the random projections in use, $f(r) - f(q)$
has the same distribution as that of a point at distance
$\sqrt(d(t,q)^2+1/c^2)$ from $q$ -- clearly this distance is greater than
$1$.

\begin{remark}
Although the converse of lemma \ref{couponcoll} is not always true -- that
is,
it is not necessary that $2^{I(X)}$ random samples for a required to guess
the value of
a random variable --
it can be shown that for the specific hash functions in consideration this
is the case.
That is, we need $2^{(1\pm o(1))kM/\log(1/g)}$ random samples to
guess the
value of $H(p)$
given $F(q)$. The essential idea is to consider different values of $f(q)$
in small
increments of $\eps/c$ and argue that the number of projections for which
$R(q)$
lie in a small interval is close to the expected value with high
probability and then
argue that we need close to the corresponding number of guesses for those
set of intervals.
\end{remark}

\section{Implementation Discussion}\label{secimplementation}

We may assume that $d$ is at most $O(\log n)$ as for larger $d$ we can use dimension reduction
 techniques that preserve distances. 
Alternately we may use 
$O(\log n)$ locality preserving hash functions to represent a point in the database.
So we need not store the entire description of each point
in the hash table but only its $O(\log n)$ size hash value.  
This makes the size of each hash entry small especially if we know that there is a 
unique $(r,cr)$-approximate nearest neighbor. More succinct representations that use
close to at most $2 \log n$ bits can be obtained by first embedding the points into a
high-dimensional hamming metric  and then reducing the number of dimensions
to about $O(\log n)$ by XORing suitable sized random subsets of the bits (see lemma 1 in \cite{KOR98}).
If the nearest neighbor
is unique then in the final representation, each bit of the query and the nearest neighbor
will differ with probability at most $1/(2c)$ whereas for other neighbors each bit position
will differ with probability at least $\frac{1}{2}(1-1/e)$. A simple and tight probability calculation shows
that for large enough $c$, $2 \log n$ bits suffice with high probability to distinguish 
the nearest neighbor from the other points. Note that the hash key $H(p)$ need not be
stored explicitly as it suffices to hash this into an index for the hash array.

While we have included several $polylog n$ factors in the space complexity
these are unlikely to be
required in practice.  The first $\log n$ factor comes from Lemma
\ref{couponcoll} and is required only for
arbitrary random variables.  In our case since the entropy is obtained by
adding several different independent
random variables it is easy to show that this is not required.  The second
 $\log n$ comes from the application of Markov
inequality on the entropy distribution.  This again can be
eliminated by using say Chebyshev or
Chernoff bounds.  The third one arises by the crude application of Markov's
inequality to ensure that not too many
far off points are examined in each hash table. Again we expect this will
not be really required in practice.
So for large enough constant $c$, if we are searching for a unique $(r, cr)$ nearest
neighbor, the total amount of space required in
practice is close to $2n\log n$ bits.
Even for $n$ equal to a million, this is the only 5MB which is a tiny
fraction of the main memory space available
on PC's.


\section {Appendix}

\subsection {$I(h(p)|f(q))$ for Random Instance}\label{entropyqp}

We will show that $I(h(p)|f(q)) = I(h(q)|f(p))$ for the random instance 
of nearest neighbor search in Euclidean space presented in section \ref{secrandom}.
$p$ is a random point distributed as $N^d(0,1/\sqrt 2)$, and $q$ is distributed as 
$N^d(p,1/c)$. We will compute the probability density that $f(q)=y$, and conditioned on 
this the probability that $h(p)\not = h(q)$.

After the random projection, $f(q)$ is distributed as $N(0,\sqrt (1/2+1/c^2))$.
Also, the probability density function of $f(p)$ conditioned on $f(q)=y$ is 
$Pr[f(p) = x] Pr[f(q) = y|f(p) = x]/Pr[f(q) = y]
= \eta(x \sqrt 2) \eta((x-y)c)/\eta(y/\sqrt (1/2+1/c^2))$, which is $\eta(c^2 y/(2+c^2), 1/\sqrt (2+c^2))$, 
the normal distribution with mean $c^2 y/(2+c^2)$ and deviation $1/\sqrt (2+c^2)$.

So given that $f(q)=y$, probability that $h(p)\not = h(q)$ is 
$\Phi(\sqrt (2+c^2) c^2 y/(2+c^2)) = \Phi( c y/\sqrt (1+2/c^2))$. Since the probability density function
of $f(q)$ is also given by $\eta(\sqrt 2 y/\sqrt (1+2/c^2))$, this results in the same calculation as
for $I(h(p)|f(q))$ except that the variables are scaled by a factor of $\sqrt (1+2/c^2)$.
So, $I(h(q)|f(p)) = I(h(p)|f(q))$.

\subsection {Thick Hyperplanes}\label{thickhyper}
Let us consider the case when $d$ is very large say at least $n \log n$. 
In that case we choose special hyperplanes that better separate the set of points. The 
hyperplanes are obtained as follows.

If $v_1, .., v_n$ denote the points of the database, choose $a_i$ randomly to be either $+1$ or $-1$
and set $h = \sum a_i v_i$. 
Observe that $h$ is a random variable from $N^d(0,\sqrt d)$.
So if $p$ is random point in $B(q,r)$ then $h.p-h.q$ is distributed as $N(0,r)$
We will show that $h$ separates the set of points well. 
Indeed, look at $h.v_i = a_i |v_i|^2 + \sum a_j v_i.v_j$.
Note that $v_i.v_j$ is very small, distributed as $N(0,1/\sqrt d)$. 
So the sum is distributed as $N(0, \sqrt(n/d))$. $|v_i|^2$ is concentrated around $1$
and at least $1- \eps$ with high probability (at least $1-exp(-O(d))$). Also
the sum term is at most $\eps$ with probability at least $1-exp(O(\log n))$.
So with high probability over the $\log n$ projections for all such $h,
|h.v_i| > (1-\eps)$. 
Now, since the probability that $h.q\not = h.p$ is clearly at most $exp(-O(c^2))$,
we have $I (h(q)/f(p))$ is $O(c^2 exp(-O(c^2)))$.

This argument can also be applied if $d$ is as small as $n$ but deriving the appropriate hyperplanes may 
require solving a system of equations. If $a$ is the column vector with entries as $a_i$, then 
$h$ is obtained by solving $Ah=a$, where the rows of $A$ are the point vectors $v_i$. 

\end{document}